# Nano-Optical Device Design with the Use of Open-Source Parallel Version FDTD and Commercial Finite Element Package

Y. Liu
Department of Electrical and Computer Engineering
Texas A&M University
College Station, Texas, USA
ylgogogo@tamu.edu

K. Chang
Department of Electrical and Computer Engineering
Texas A&M University
College Station, Texas, USA
chang@ece.tamu.edu

*Abstract*—In this paper, the implementation of open-source parallel-version FDTD (Finite-Difference-Time-Domain) software, MEEP, on Texas A&M supercomputers and commercial finite element package, COMSOL, on a single workstation for the design design of nano-optical device is reported. The the computer architecture and performance of both numerical methods on the same design will be briefly described.

## I. INTRODUCTION

Due to the growing research in nano technology, the requirement of numerical tools becomes demanding for accurate and fast design. The shrinking electronic or optical devices have brought intense computational burden than ever for the accuracy-concerned refined computational meshes. Based on the above reason, high performance computers are needed to achieve multiple design tasks. At present, there are couple commercial electromagnetic simulators available for the design of nano or THz devices. However, the general public prefer open-source softwares for cost, license, code modification and sharing concerns. In this paper, the implementation of an open parallel Finite-Difference Time-Domain (FDTD) software, MEEP, on TAMU supercomputer Eos and commercial package, COMSOL, based on Finite Element Method (FEM) on a single workstation is reported for the design of nano-optical devices. Due to the difference of numerical algorithm between FDTD and FEM for the material interface treatment, FEM obviously works better than FDTD for the designed problem.

## II. MEEP AND COMSL

MEEP (MIT-Electromagnetic Equation Propagation) is a free finite-difference time-domain package developed by MIT since 2006 for computational electrodynamics. Currently, it can support both 2-D and 3-D simulation for various physics problems with either serial or parallel computation mode. Due to its powerful functionality and open-source nature, MEEP now is well-known in optics society and is intensely used in the area of optics and electromagnetics. The code is written in C++ and can be performed on most Unix/Linux based operating systems. Various prerequisite packages have to be installed in advance before porting MEEP to the system. For more details, the information of this package can be found on MEEP's website [1].

COMSOL is a Finite Element based package that can provide both 2D and 3D multiphysics simulation [2]. Currently, it has packages for electrical, mechanical, fluid and chemical related simulation. Compared with another FEM [5] package, HFSS, COMSOL gives users the freedom to edit their desired equations that the built-in library doesn't have, which is a great incentive for many customers.

## III. FEATURES OF SIMULATION MACHINES

MEEP and COMSOL are installed on TAMU Eos [3] and a Dell Precision 690 workstation respectively. For better understanding about the machines, Table I lists some important features of Eos and Dell Precision 690.

TABLE I. FEASTURES OF DELL WORKSTATION AND EOS

|  | **Dell Precision 690** | **TAMU Eos** |
| --- | --- | --- |
| **OS** | Linux | Linux |
| **Number of nodes** | 1 | 372 |
| **Cores** | 4 | 8 (majority) or 12 / per node |
| **RAM** | 32GB | 24GB / per node |
| **CPU** | Intel Xeon 5160 Dual-core 64bit @3GHz | Intel Nehalem quad-core X5550 64-bit@2.8GHz |
| **Vendor** | Dell | IBM (iDataPlex) |
| **Storage** | 147GB | 9,056GB |
| **Peak Performance** | NA | 35.5TFlops |
| **Interconnect** | None | InfiniBand. Bandwidth : 4GB/sec point-to-point |
| **MPI used** | None | OpenMPI 1.4.3 |

## IV. SIMULATION OF NANO-OPTICAL DEVICE

Fig. 1. and Fig. 2. shows H field of the designed device simulated by MEEP and COMSOL respectively.

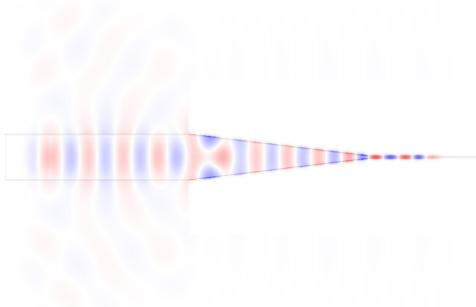

Fig. 1. Simulation result of MEEP.

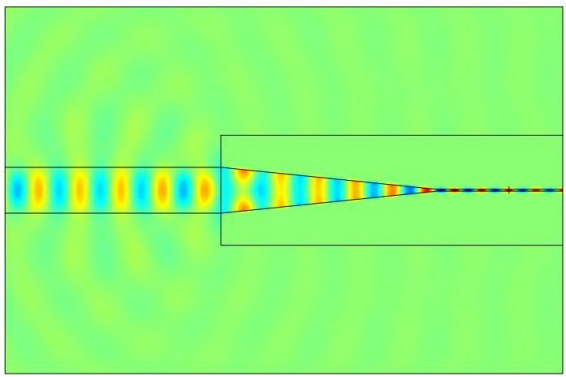

Fig. 2. Simulation result of COMSOL

For MEEP, the dimension of the computational window is 15.5 x 10 (all physical parameters are normalized so no unit noted here) and the total number of grid is 6,200,000. Due to the nature of FDTD [4], long-time iteration is required to achieve numerical stability and accuracy. Table II lists the wall clock time Eos took for each submitted job with different core number requested. Based on this statistics, the simulation time cannot be reduced further with more than 96 total core number.

TABLE II. RUN TIME FOR DIFFERENT CORE SPECIFICATION ON EOS

(TIME FORMAT -> HOUR:MINUTE:SECOND)

| Total Core Number | 8 Tasks/ node |
|---|---|
| 8 | 7hrs:09:48 |
| 16 | 10hrs:48:02 |
| 32 | 4hrs:51:10 |
| 48 | 1hr :28:48 |
| 64 | 1hr :09:53 |
| 80 | 1hr :02:16 |
| 96 | 00hr:48:19 |
| 128 | 00hr:44:51 |
| 144 | 00hr:43:14 |

On the other hand, for the same problem, the simulation widow of COMSOL is also 15.5 x 10. However, the number of unknown used here is 159,512 to achieve required accuracy and stability and the computation time is less than 1 minute.

## V. CONCLUSION

In this literature, the implementation of MEEP (FDTD) on TAMU Eos and COMSOL (FEM) on a single Dell workstation for the design of nano-optical devices is reported. Though current state-of-art high performance computers are powerful to deal with challenging problems, the choice of numerical method and analysis model to save the expensive computational resources are still important for more cost-effective development of nanotechnology in the near future.

## ACKNOWLEDGMENT

The author would like to acknowledge Supercomputing Center of Texas A&M University for providing computing resources; Mr. Fwu from SGI, Mr. Thompson and Mr. Gray from bp for providing workstations.